\documentclass{article}

\usepackage[nonatbib, final]{neurips_2022}

\usepackage[utf8]{inputenc} 
\usepackage[T1]{fontenc}    
\usepackage{hyperref}       
\usepackage{url}            
\usepackage{booktabs}       
\usepackage{amsfonts}       
\usepackage{nicefrac}       
\usepackage{microtype}      
\usepackage{xcolor}         

\usepackage{todonotes}
\usepackage{amsmath}
\usepackage{graphicx}
\usepackage{float}
\usepackage{algorithm}
\usepackage{algorithmic}
\usepackage{bbm}

\DeclareMathOperator*{\argmin}{arg\,min}
\newcommand{\equptoadd}{\ensuremath \stackrel{+}{=}}
\newcommand{\qphi}{ q}

\newtheorem{theorem}{Theorem}[section]
\newtheorem{lemma}[theorem]{Lemma}

\definecolor{tab:blue}{RGB}{32,121,178}
\definecolor{tab:green}{RGB}{51,158,51}
\definecolor{tab:orange}{RGB}{254,125,33}
\definecolor{tab:gray}{RGB}{127,127,127}
\definecolor{tab:purple}{RGB}{147,106,187}

\title{VaiPhy: a Variational Inference Based Algorithm for Phylogeny}

\author{%
Hazal Koptagel$^{12*}$ \quad Oskar Kviman$^{12*}$ \quad Harald Melin$^{12}$ \\ 
\textbf{Negar Safinianaini}$^{12}$ \quad \textbf{Jens Lagergren}$^{12}$ \\
$^1$School of EECS, KTH Royal Institute of Technology, Stockholm, Sweden \\
$^2$Science for Life Laboratory, Solna, Sweden \\
$^*$Equal contribution \\
\texttt{\{koptagel,okviman,haraldme,negars,jensl\}@kth.se}\\
}

\begin{document}

\maketitle

\begin{abstract}
Phylogenetics is a classical methodology in computational biology that today has become highly relevant for medical investigation of single-cell data, e.g., in the context of cancer development. 
The exponential size of the tree space is, unfortunately, a substantial obstacle for Bayesian phylogenetic inference using Markov chain Monte Carlo based methods since these rely on local operations. And although more recent variational inference (VI) based methods offer speed improvements, they rely on expensive auto-differentiation operations for learning the variational parameters.
We propose VaiPhy, a remarkably fast VI based algorithm for approximate posterior inference in an \textit{augmented tree space}. VaiPhy produces marginal log-likelihood estimates on par with the state-of-the-art methods on real data and is considerably faster since it does not require auto-differentiation. Instead, VaiPhy combines coordinate ascent update equations with two novel sampling schemes: 
(i) \textit{SLANTIS}, a proposal distribution for tree topologies in the augmented tree space, and (ii) the \textit{JC sampler}, to the best of our knowledge, the first-ever scheme for sampling branch lengths directly from the popular Jukes-Cantor model. We compare VaiPhy in terms of density estimation and runtime. Additionally, we evaluate the reproducibility of the baselines. We provide our code on GitHub: \url{https://github.com/Lagergren-Lab/VaiPhy}. 
\end{abstract}

\section{Introduction}
Phylogenetic software has for a long time been applied in biological research, and biological findings relying on phylogenetic trees are frequent. Moreover, due to the emergence of single-cell sequencing, phylogenetic inference for bifurcating and multifurcating trees can now be utilized in medical studies of how cells differentiate during development and how tumor cells progress in cancer. 

When attempting to infer the posterior distribution over phylogenetic trees for $|X|$ taxa, represented as sequences, the main obstacle is the exponential size of the tree space, i.e., the set of candidate trees having those taxa as leaves. In Markov chain Monte Carlo (MCMC) based Bayesian phylogenetic inference, e.g., MrBayes \cite{ronquist2012mrbayes} and RevBayes \cite{hohna2016revbayes}, this tree space is explored by performing local operations on the trees. In order to avoid local operations, Bouchard-C\^{o}t\'{e} et al. \cite{BouchardCote:2012ie} launched Sequential Monte Carlo (SMC) as an alternative methodology for Bayesian phylogenetic inference.

Probabilistic phylogenetic inference is a machine learning problem, and the key to unlocking this crucial application can be expected to be found in the machine learning methodology; in particular,  considering the capacity of the VI methodology, \cite{Wainwright:2007du,blei2017variational}, to deliver impressive performance gains for Bayesian inference, e.g., compared to MCMC. Expectedly, such gains have recently been made within the realm of phylogenetic analysis.

For instance, inference based on the computationally costly CAT model, which allows different equilibrium frequencies across sequence sites, has been made substantially more efficient by the application of VI, \cite{10.1093/molbev/msz020.Dang}.
Also, \cite{10.7717/peerj.8272.Fourment} applied VI to render the analysis of a sequence evolution model more efficient. They consider the efficiency improvements obtained by VI, in particular using the generic STAN software \cite{10.7717/peerj.8272.Fourment}, when analyzing a single fixed tree topology. Interestingly, the VI approach can also be used to analyze a large number of candidate phylogenetic trees. In \cite{zhang2018generalizing},
so-called subsplit Bayesian networks (SBNs) were introduced in order to represent the posterior over a given set of candidate trees. This approach has subsequently been developed further by introducing an edge (a.k.a. branch) length distribution \cite{zhang2018variational} and making the edge length distribution more expressive by taking advantage of normalizing flows \cite{Zhang0if}. In particular, \cite{zhang2018variational} obtains posterior distributions that improve on the marginal likelihood obtained by using MrBayes, \cite{huelsenbeck2001mrbayes}, in combination with the improved Stepping Stone (SS) method, \cite{10.1093/molbev/msq224.Fan}.

The SBN approach evades the exponentially sized tree space by considering a given set of candidate trees and the partitions of the given taxa defined by those candidate trees. 
 It is noteworthy that the VI methodology earlier has been successfully applied to bipartite matching and a few other combinatorial problems \cite{BouchardCote:VC8xIIC9}; the phylogeny problem, however, involves both combinatorial and continuous parameters. 
 
 Multifurcating trees are common in biology, and biologists are well familiar with them. For instance, in a clone tree that represents the evolutionary relationships between the subclones of a tumor, nodes may have more than two children. Even in cases with no biological support for multifurcating trees, biologists are used to working with multifurcating trees. E.g., the consensus method used to analyze the posterior of a MrBayes run may produce a multifurcating tree.

\begin{figure*}[!t]
    \centering
    \includegraphics[width=12cm]{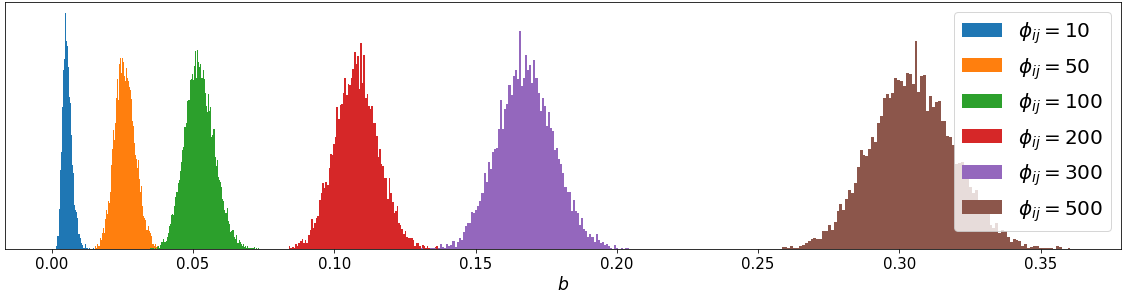}
    \caption{Visualization of branch lengths, $b$, sampled directly from the Jukes-Cantor model. The figure shows samples for different numbers of estimated mutations along an edge, $\phi_{i,j}$, when the total number of sites are $M=2000$.}
    \label{fig:jc_sampler}
\end{figure*}

{\bf Contributions:}
We present a novel framework for phylogenetic inference that produces marginal likelihood estimates on par with the state-of-the-art VI methods while being up to 25 times faster than VBPI-NF on real datasets. The framework is on the top level a mean-field VI algorithm and is comprised of the following three novel methods: 
\begin{itemize}
    
    \item \textbf{SLANTIS} is a proposal distribution that proposes multifurcating spanning trees.  Sampling from SLANTIS is performed following an approach suggested by \cite{Diaconis} for sampling a perfect matching in bipartite graphs: it is fast, and the sampled tree's compensation weight is easily computed without additional cost of time.
    \item \textbf{The JC sampler} is a novel scheme for sampling branch lengths directly from the Jukes-Cantor (JC69) model \cite{jukes1969evolution}. This is a significant contribution, as no previous sampling scheme is known for this task and the JC model arguably is a commonly used substitution model in phylogenetics.
    \item \textbf{VaiPhy} is a VI algorithm for obtaining posterior approximations over the set of branch lengths, the ancestral sequences, and tree topologies, given the leaf set observations. VaiPhy utilizes SLANTIS and the JC sampler in order to compute the importance weighted evidence lower bound (IWELBO; \cite{Burda}) and its update equations. Compared to existing VI methods in phylogenetics, VaiPhy extends the target distribution by considering an auxiliary variable, the ancestral sequences, and it does not require any input consisting of precomputed trees.
\end{itemize}
We demonstrate the flexibility of our framework by using its learned parameters to parameterize two new proposal distributions for the Combinatorial SMC (CSMC) algorithm. We compare our CSMC method, $\phi$-CSMC, with the other two main SMC based methods, outperforming them in terms of marginal likelihood estimation on real data.

Finally, we contribute to the field of Bayesian phylogenetic inference by benchmarking the most prominent methods in the field on seven popular real datasets.

\section{Background} 
In the Bayesian phylogenetic setting, we are interested in the posterior distribution over phylogenies -- pairs of tree topologies and sets of branch lengths -- given observations on the leaves,
\begin{equation}
\label{eq:vanilla model}
    p_{\theta}(\tau, \mathcal{B}|X) = \frac{p_{\theta}(\tau, \mathcal{B}, X)}{p_{\theta}(X)},
\end{equation}
where $\tau$, $\mathcal{B}$, and $X$ denote the tree topology, the set of branch lengths and the observations, respectively. More specifically, $\mathcal{B}$ contains a single branch length per edge, $e$, in the set of edges in $\tau$, i.e., $\mathcal{B}=\{b(e) : e\in E(\tau)\}$. Unfortunately, $p_{\theta}(\tau, \mathcal{B}|X)$ is intractable,  so we need to approximate it using Bayesian inference. Before going into how to approximate this quantity, we first provide some crucial notation.

In phylogenetics, one is often concerned with bifurcating $X$-trees. These trees have labeled leaf nodes, and their internal vertices have either two (root in a rooted tree) or three degrees. Meanwhile, multifurcating trees' internal vertices can have higher degrees. We use $M$, $\Sigma$, and  $N$ to denote the sequence length, the set of nucleotides, and the number of vertices, respectively. Further, we let $|X|$ be the number of given taxa, $V$ and $E$ be the set of vertices and edges, while $(V, E)$ is an acyclic graph.

Let $Z_i^m$ be the probabilistic ancestral sequence of node $i$ at position $m$, and $\theta$ denote the model parameters, such as the branch length prior parameters or the parameter choices in the substitution model. Then, the evolutionary model along an edge, $e = (i,j)$, is $ p_{\theta} (Z_i^m = \alpha| Z_j^m = \beta) = exp( b(e)\; Q_{\alpha \beta} ) $, where $Q$ is the rate matrix, parameterized according to the JC69 model, and $\alpha, \beta \in \Sigma$ are nucleotides. The  likelihood is defined as 

$$p_\theta(X|\tau, \mathcal{B})= \sum_{Z\in \Sigma^M }p_\theta(X, Z|\tau, \mathcal{B}),$$
and can be computed effectively using Felsenstein's pruning algorithm \cite{felsenstein1981evolutionary}.

\section{Proposed Framework}
Existing Bayesian phylogenetic approaches often perform inference in a sample space over bifurcating $X$-trees and branch length sets, $\Omega$. In contrast, we propose to work in an augmented sample space, $\mathrm{A}$. Firstly, we work in a tree topology space for multifurcating trees with labeled internal vertices. Secondly, we extend the target distribution in Eq. \eqref{eq:vanilla model} by considering ancestral sequences on the internal vertices as an auxiliary variable, $Z$. 

Apart from that, our augmented space offers higher modeling flexibility (clearly, since $\Omega \subset \mathrm{A}$); it enables posterior estimation over $Z$, which further lets us estimate the number of mutations between all vertices in the sample space. This is a key insight that ultimately allows us to devise VaiPhy, a mean-field coordinate ascent VI \cite{hoffman2013stochastic}  algorithm. 

\subsection{VaiPhy}

In the augmented space, $\mathrm{A}$, we wish to approximate the posterior distribution
\begin{equation}
    p_\theta(\tau, \mathcal{B}, Z|X) \propto p_\theta(X, Z |\mathcal{B}, \tau)p_\theta(\mathcal{B} |\tau) p_\theta(\tau),
\end{equation}
using a factorized variational distribution

\begin{equation}
\label{eq:factorized q}
    \qphi(\tau, \mathcal{B}, Z|X) = \qphi( \mathcal{B} |X)\qphi(\tau|X)q(Z|X), 
\end{equation}

where $q(Z|X) = \prod_{i\in I(\mathrm{A})}q(Z_i|X)$. Here, $I(\mathrm{A})$ denotes the set of internal vertices in the augmented space. We will use $q(Z|X)$ to compute
$\phi_{ij}\in [0, M]$, the expected number of mutations over the edge $(i, j)$, for all possible edges. Notice that the maximum number of observed mutations between two vertices is the number of sites, $M$.

As is standard in classical VI, we alternatively update the distributions in Eq. \eqref{eq:factorized q} in order to maximize the evidence lower bound (ELBO),
\begin{equation}
    \label{eq:elbo}
    \mathcal{L} = \mathbb{E}_{\qphi(\tau, \mathcal{B}, Z|X)}\left[ \log \frac{p_\theta(X, Z |\mathcal{B}, \tau)p_\theta(\mathcal{B} |\tau) p_\theta(\tau)}{\qphi( \mathcal{B} |X)\qphi(\tau|X)q(Z|X)} \right],
\end{equation}
over iterations. Each iteration, we also update $\phi$, following \cite{Friedman:2002bo}
\begin{equation}
\label{eq:phi update}
    \phi_{ij} =
    \sum_{\alpha\neq \beta}F^{\alpha, \beta}_{ij},
\end{equation}
where $F^{\alpha,\beta}_{ij}$ is the estimated number of times a nucleotide mutates from $\alpha$ to $\beta$, defined as
$
    F^{\alpha, \beta}_{ij} = \sum_{m=1}^M q(Z^m_i=\alpha|X^m)q(Z^m_j=\beta|X^m).
$

The update equations for the variational distributions are derived in a mean-field fashion. Starting with $q(Z_i|X)$, we find that
\begin{equation}
    \label{eq:upd z}
    \begin{aligned}
    \log q^*(Z_i|X) &\equptoadd \mathbb{E}_{\qphi( \tau, \mathcal{B}, Z_{\neg i}|X)}\left[
    \log p_\theta(X, Z, \mathcal{B}, \tau)
    \right] \\
    &\equptoadd \mathbb{E}_{\qphi(\tau, \mathcal{B}| X)} \left[ \sum_{Y_j \in Y_{N_\tau(i)}} q(Y_j|X) \log p_\theta(Y_j | Z_i, b(i,j))  \right], 
    \end{aligned}
\end{equation}

where  $\equptoadd$ denotes equal up to an additive constant, $Y_{N_\tau(i)}$ is the set of sequences of node $i$'s neighbors in $\tau$, and $\neg i$ is the set of latent nodes except node $i$. Continuing with the two other approximations, we have
\begin{equation}
    \label{eq: upd tau}
    \begin{aligned}
    \log q^*(\tau|X) &\equptoadd \mathbb{E}_{\qphi(\mathcal{B}, Z|X)}\left[
    \log p_\theta(X, Z, \mathcal{B}, \tau)
    \right] \\
    &\equptoadd \mathbb{E}_{\qphi(\mathcal{B}|X)}   \left[  
    \sum_{{(i,j) \in E(\tau) \atop  Z_i, Z_j}} q(Z_i, Z_j|X) \log p_\theta(Z_i | Z_j, b(i,j))
    \right] 
    \end{aligned}
\end{equation}

and

\begin{equation}
    \label{eq: upd B}
    \begin{aligned}
    \log q^*(b(i,j)|X) &\equptoadd \mathbb{E}_{\qphi(\tau, \mathcal{B}_{\neg b(i,j)}, Z|X)}\left[
    \log p_\theta(X, Z, \mathcal{B}, \tau)
    \right]\\ 
    &\equptoadd \mathbb{E}_{\qphi(\tau|X)} \left[ \sum_{Z_i, Z_j}  \qphi(Z_i, Z_j|X) \log p_\theta(Z_i | Z_j, b(i,j), \tau) \right] + \log p_\theta(b(i,j)).
    \end{aligned}
\end{equation}
    
The exact computation of the expectations in Eq. \eqref{eq:upd z} and \eqref{eq: upd B} involves summing over all possible trees, which is computationally intractable. Luckily, we can easily approximate the expectation using importance sampling, and our new proposal distribution, $s_\phi(\tau)$, is outlined in the subsequent Sec. \ref{sec:slantis}. We direct the reader to Appendix~\ref{appendix:vaiphy_general} for the derivations of the above variational distributions and the details of the algorithm.

To obtain good starting points for the variational distributions, we initialize $\phi$ with trees obtained using the Neighbor-Joining algorithm \cite{gascuel1997bionj}. The details of the initialization are provided in Appendix~\ref{appendix:nj_init}.
Alg. \ref{alg:vaiphy} is a high-level algorithmic description of VaiPhy. The proof regarding the natural gradient is presented in Appendix~\ref{appendix:ng_proof}.

\begin{algorithm}[H]
\caption{VaiPhy in pseudocode -- $\iota$ is the total number of iterations and $\eta$ is the natural gradient step size}
\label{alg:vaiphy}
\begin{algorithmic}[1]
\STATE {\bfseries Input:} $X, \phi, \eta, \iota$
   \FOR{$\text{iter}=1,\ldots,\iota$}
   \STATE Approximate $\log q^*(Z|X)$ using $s_\phi(\tau)$ \COMMENT{Eq. \eqref{eq:upd z}}
   \STATE $q(Z_i)\leftarrow  (1-\eta)q(Z_i) + \eta q^*(Z_i)$ 
   \STATE $\phi_{ij}\leftarrow \sum_{\alpha\neq\beta}F^{\alpha,\beta}_{ij}$ \;\;  \COMMENT{Eq. \eqref{eq:phi update}}
   \STATE $\qphi(\tau|X) \leftarrow q^*(\tau|X)$ \;\COMMENT{Eq. \eqref{eq: upd tau}}
   \STATE $\qphi(\mathcal{B}|X) \leftarrow q^*(\mathcal{B}|X)$ using $s_\phi(\tau)$ \COMMENT{Eq. \eqref{eq: upd B}}
   \ENDFOR
   \STATE \textbf{return} $\qphi(\tau, \mathcal{B}, Z|X)$, $\phi$
\end{algorithmic} 
\end{algorithm}

\subsection{SLANTIS} 
\label{sec:slantis}
We propose the method Sequential Look-Ahead Non-local Tree Importance Sampling (SLANTIS), which enables sampling of a tree, $\tau$, from a distribution, $s_\phi(\tau)$. SLANTIS has support over multifurcating trees with labeled internal vertices and is parameterized by $\phi$. To describe the algorithm, we next introduce some notation. 

Using $\phi$, SLANTIS immediately constructs $W$, an $N\times N$ matrix describing all pairwise edge weights based on the substitution model,
\begin{equation}
\label{eq:weight matrix}
    W_{ij} = \sum_{\alpha, \beta:\alpha\neq\beta} \phi_{ij} \; (b(i, j) \; Q_{\alpha\beta}) + \sum_{\alpha, \beta:\alpha=\beta} (M - \phi_{ij}) \; (b(i, j) \; Q_{\alpha\beta}),
\end{equation}
where $Q$ is the rate matrix of the JC69 model and $\alpha$ and $\beta$ are nucleotides.
Moreover, let $\mathcal{G}$ be a graph with vertex set $\{N\}$ with edge weights specified by $W$. Let $\mathcal{G_I} = \mathcal{G} \text{\textbackslash} \Lambda$, where $\Lambda$ is the $(V,E)$ of the leaves, i.e., the graph induced by the internal vertices of our trees. The pseudocode of the algorithm is presented in Appendix \ref{appendix:slantis}. The main idea, which  is borrowed from \cite{Diaconis}, is to order the edges of the graph and for each edge in the order, make a choice whether to include it in the sampled tree or not. The choice is made based on a Bernoulli trial, where the two probabilities are proportional to the weight of the maximum spanning tree (MST) containing the  already selected edges and the currently considered edge and that of the MST containing the  already selected edges but not the currently considered edge.\footnote{In \cite{Diaconis}, the candidate edges are selected from those that will make a \textit{perfect matching}, and the probability of each edge is uniform.} The action associated with the former outcome is to include the edge in the sampled tree, and the other outcome is not to include it.

\begin{figure}[b]
\begin{center}
\includegraphics[width=\textwidth]{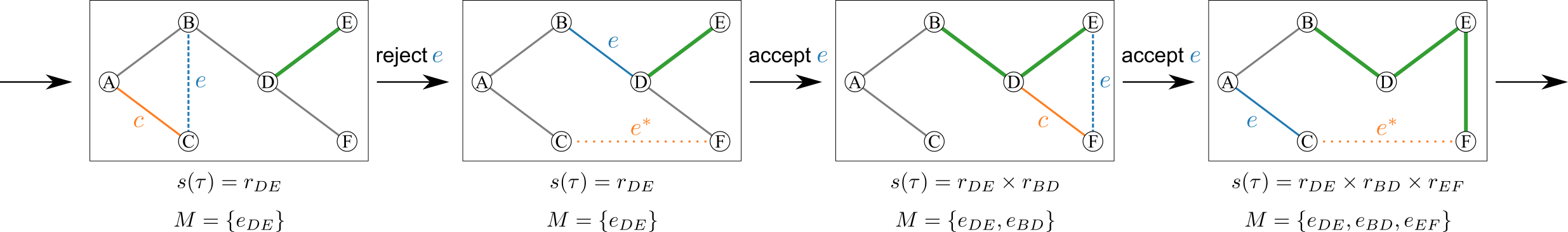} 
\caption{Propagation step of SLANTIS. The solid line is $\tau$ so far, and bold \textbf{\textcolor{tab:green}{green}} lines are accepted edges. Dashed and dotted lines are the edges not in $\tau$. The \textcolor{tab:blue}{blue} color indicates the edge we consider 
($e$),
and the \textcolor{tab:orange}{orange} color indicates the alternative edge, either
$c$ or $e^*$.
$r_e = W(e) / (W(e) + W(c))$ or $r_e = W(e) / (W(e) + W(e^*))$. At the end of SLANTIS, the spanning tree with bold green lines is reported.}
\label{fig:slantis}
\end{center}
\vskip -0.2in
\end{figure}

To facilitate fast computation of the required MSTs, we first consider the edges of an MST $T_1$ of $\mathcal{G}$.
After the edges of $T_1$, the edges of an MST $T_2$ of $\mathcal{G}\setminus  T_1$ can be found. In general, $T_s$ is an MST of $\mathcal{G}\setminus \cup_{r=1}^{s-1} T_r$, and  after the edges of $\cup_{r=1}^{s-1} T_r$, the edges of $T_s$ can be found. In the algorithm, we repeatedly take advantage of the property that for any tree $T$ and edge $e$ not in $T$,  $T\cup e$ contains a unique cycle, and the removal of its heaviest edge different from $e$ produces the minimum weight sub-tree  of  $T\cup e$  containing $e$. If the considered edge is in $T$ and we choose to include it in the sampled tree, $\tau$, we recursively update $\log s_\phi(\tau)$ by adding the Bernoulli log probability corresponding to the edge. See Fig. \ref{fig:slantis} for an illustration.

\subsection{The JC Sampler}
The JC69 model is arguably one of the most frequently used substitution models in Bayesian phylogenetics. However, as far as we are aware, the model has only been used for obtaining maximum likelihood estimates of branch lengths. Here, we observe that the JC69 model can be seen as an i.i.d. Bernoulli likelihood over sites and leverage the Beta-Bernoulli conjugacy for obtaining a normalized distribution over branch lengths in variational inference. Below we show how to sample from this distribution and evaluate its likelihood using the change-of-variables formula.

Given a branch length $b(e)=b$, the probability of the $m$'th nucleotide \textit{not} mutating along edge $e=(i,j)$ is, according to the JC69 model,
\begin{equation}
\label{eq:jc p}
    p = \left(\frac{1}{4} + \frac{3}{4}e^{-\frac{4}{3}b}\right)^{\mathbbm{1}\{Z_i^m=Z_j^m\}}.
\end{equation}
As all mutations are equiprobable for this model, the probability of the nucleotide mutating is thus $1-p$,
giving rise to the following distribution when all $M$ estimated mutations and non-mutations are observed and assuming mutations occur independently across sites 
\begin{equation}
\begin{aligned}
    \text{Bern}\left( \phi_{ij} | p \right) &= \prod_{m=1}^M \text{Bern} \left( \phi_{ij}^m | p \right) = \prod_{m=1}^M p^{1-\phi_{ij}^m} (1-p)^{\phi_{ij}^m} = p^{M-\phi_{ij}}(1-p)^{\phi_{ij}},
\end{aligned}
\end{equation}
where $\phi_{ij}^m$ is the expected number of mutations over the edge $(i,j)$ at site $m$.
Assuming an uninformative Beta prior on the Bernoulli parameter, $p$, and recalling that Beta is a conjugate prior for the Bernoulli distribution, we get a posterior distribution over $p$,
    $\text{Beta}(p| M-\phi_{ij} + 1, \phi_{ij} + 1)$. Since we are ultimately interested in sampling $b$, not $p$, we find a transform, $f(\cdot)$, mapping $p$ to $b$, using Eq. \eqref{eq:jc p};

\begin{equation}
    f^{-1}\left(b\right) = 
    \frac{1}{4} + \frac{3}{4}e^{-\frac{4}{3}b}
     = p
\end{equation}
\begin{equation}
\label{eq:f}
    f(p) = -\frac{3}{4}\log\left( \frac{4}{3}\left(p - \frac{1}{4}\right)\right) = b.
\end{equation}
In all, sampling branch lengths from the JC model include two straightforward steps, (i) draw $p$ from the Beta posterior distribution, and (ii) pass it through the transform, $b=f(p)$.

Of course, we are also interested in evaluating the likelihood of the sampled branch length. To do so, we use the
change-of-variables formula; we let
\begin{align}
\label{eq:change-of-variable}
    s_\phi(b) &= \text{Beta}(p| M-\phi_{ij} + 1, \phi_{ij} + 1)\left\vert\frac{df^{-1}(b)}{db} \right\vert\\
    &= \frac{p^{M-\phi_{ij}} (1-p)^{\phi_{ij}}e^{-\frac{4}{3}b}  }{\text{B}(M-\phi_{ij} + 1, \phi_{ij} + 1)},
\end{align}
where B$(\cdot)$ is the Beta function. This provides an efficient way of evaluating the likelihood of a branch length. In Alg. \ref{alg:jcsampler}, we summarize the JC sampler with an algorithmic description.

\begin{algorithm}[H]
\caption{The JC Sampler}
\label{alg:jcsampler}
\begin{algorithmic}[1]
\STATE {\bfseries Input:} $\phi, i,j$
   \STATE Sample $p\sim \text{Beta}(p| M-\phi_{ij} + 1, \phi_{ij} + 1)$
   \STATE Transform $b = f(p)$ \; \COMMENT{Eq. \eqref{eq:f}}
   \STATE Evaluate $s_\phi(b)$ \; \COMMENT{Eq. \eqref{eq:change-of-variable}}
   \STATE \textbf{return} $b, s_{\phi}(b)$
\end{algorithmic} 
\end{algorithm}

\subsection{Assessing the Framework}
The framework learns its parameters and distributions by maximizing the ELBO in Eq. \eqref{eq:elbo}. However, the IWELBO offers a tighter lower bound on the marginal log-likelihood. Using SLANTIS and the JC sampler, we can easily evaluate our framework using the IWELBO:
\begin{equation}
    \label{eq:iwelbo}
    \mathcal{L}_{L} = \mathbb{E}_{\mathcal{B}_\ell, \tau_\ell \sim s_\phi(\mathcal{B},\tau)}\left[
    \log\frac{1}{L}\sum_{\ell=1}^L\frac{p_\theta(X, \mathcal{B}_\ell, \tau_\ell)}{s_\phi(\mathcal{B_\ell},\tau_\ell)}
    \right],
\end{equation}
where we may factorize the denominator as $s_\phi(\mathcal{B},\tau) = s_\phi(\mathcal{B}|\tau)s_\phi(\tau)$, and 
$s_\phi(\mathcal{B}|\tau) = \prod_{e\in E(\tau)} s_\phi(b(e))$. Note that we have marginalized out the auxiliary variable, $Z$, in the target in Eq. \eqref{eq:iwelbo}, as we do not wish to compute $\mathcal{L}_{L}$ using our variational distributions.

\section{Using VaiPhy to Parameterize the CSMC}
\label{sec:vaiphy2csmc}
As the tree topology space in $\mathrm{A}$ is considerably larger than the bifurcating $X$-tree space, obtaining competitive  log-likelihood (LL) estimates is not straightforward. Namely, the uniform prior over the tree space in $\mathrm{A}$, $
    p_\theta(\tau) = 1/(N-|X|) ^{N - 2}$,
turns into a heavily penalizing term. When compared to baselines in smaller sample spaces, this will become a considerable negative factor for VaiPhy. Compare this to the prior when considering unrooted, bifurcating $X$-trees (which is the case for VBPI-NF): $
    p_\theta'(\tau) = 1/{(2|X| - 5)!!},$
where $|X|$ is the number of taxa. As an example, for $|X|=10$ and $N=18$, $p_\theta(\tau)$ distributes mass uniformly over $281$ trillion trees, while the latter one merely considers $2$ million trees.

Inspired by \cite{moretti2021variational}, we instead seek to use VaiPhy to parameterize a CSMC algorithm, thereby obtaining a procedure for projecting our trees to the  lower-dimensional leaf-labeled tree space. Here, the prior is instead $
    p_\theta''(\tau) = 1/{(2|X| - 3)!!}.
$

Since $\phi$ is not constrained to multifurcating trees, we can utilize the estimated number of mutations between vertices learned in $\mathrm{A}$, also in the bifurcating $X$-tree space, $\Omega$. Here arises an interesting obstacle. In the CSMC algorithm, the internal vertices are \textit{unlabeled}. Hence we are effectively blindfolded and need to estimate which entry in $\phi$ to use for a given edge. We achieve this by scoring edges based on how frequently the corresponding split of the leaf-set $X$ occurs in trees sampled from $s_\phi(\tau)$. See Appendix \ref{appendix:CSMC T proposal} for more information. 

At each rank $\rho\in[1,R]$, the CSMC proposes (i) a forest, $\mathcal{T}_{\rho}$, by \textit{merging} the roots of two subtrees in $\mathcal{T}_{\rho-1}$, and (ii) a corresponding set of branch lengths, $\mathcal{B}_\rho$. This is done in parallel for $K$ \textit{particles}, resampling particles with probability proportional to their importance weights, $w^k_{\rho}$, at every rank.\footnote{For more details and background of the CSMC algorithm, we point the reader to \cite{moretti2021variational} and \cite{wang2015bayesian} for excellent descriptions.} However, the reason to take advantage of a CSMC is that it allows us to  estimate the marginal log-likelihood in the bifurcating $X$-tree space using the well-known formula,
\begin{equation}
\label{eq:logZ}
    \log p_\theta(X) \approx \log \prod_{\rho=1}^R\frac{1}{K}\sum_{k=1}^K w_\rho^k.
\end{equation}

We let $\phi$ parameterize our two novel CSMC proposal distributions, $r_\phi(\mathcal{B}_\rho|\mathcal{B}_{\rho-1}, \mathcal{T}_{\rho})$ and $r_\phi(\mathcal{T}_{\rho}|\mathcal{T}_{\rho-1})$. The former constructs a mixture of JC samplers and evaluates the proposal's likelihood in a multiple importance sampling fashion \cite{elvira2019generalized}, while the latter forms a categorical distribution over the roots in $\mathcal{T}_{\rho-1}$, parameterized by a function of the likelihoods of trees sampled by SLANTIS. The in-detail descriptions can be found in Appendix \ref{appendix:CSMC}.

\section{Experiments}
\subsection{Visualizing the JC Sampler}We conduct a toy experiment, investigating how the estimated number of mutations along an edge affects the behavior of the JC sampler. We would expect that a greater number of mutations should generate a larger branch length. This  is the case since the involved vertices are supposedly farther apart in terms of an evolutionary distance. 

As demonstrated in Fig. \ref{fig:jc_sampler}, this advantageous behavior is indeed displayed. As $\phi_{ij}$ grows, the corresponding JC sampling distribution is shifted towards larger branch lengths. Interestingly, the JC sampler distributes its mass more symmetrically than the Log-Normal or Exponential distributions, which are typically used for modeling branch length distributions \cite{wang2015bayesian, zhang2018variational,Zhang0if,moretti2021variational}.

It follows from  Eq. \eqref{eq:f} that branch lengths grow infinitely large as $p$ goes to $0.25$. This phenomenon in phylogenetics is referred to as saturation. Saturation occurs for distantly related lineages and implies that the states of the nucleotides in these ancestral sequences are mutually independent. This is not an interesting case, hence the non-exhaustive parameter search of $\phi_{ij}$ in Fig. \ref{fig:jc_sampler}.

\subsection{Density Estimation}
Here we benchmark our methods, VaiPhy, and $\phi$-CSMC, in terms of LL estimates on seven real-world datasets, which we refer to as DS1-DS7  (\cite{hedges1990tetrapod, garey1996molecular, yang2003comparison, henk2003laboulbeniopsis,lakner2008efficiency,zhang2001molecular,rossman2001molecular}; in Appendix \ref{appendix:real_datasets}, we provide additional information about the datasets). Additionally, to the extent possible, we have used the open-source code of our baselines to survey the benchmarks on these popular datasets. We only considered baselines that employ the JC69 model. Moreover, we assumed an Exp($10$) prior over branch lengths and a uniform prior over tree topologies. This was aligned with the generative models used in all the original works of the baselines.

For all methods, we use the exact parameter configurations reported in the corresponding papers or specified in their available code. We let VBPI-NF \cite{Zhang0if} and VCSMC \cite{moretti2021variational} train for the number of iterations as specified in their works. For MrBayes \cite{ronquist2012mrbayes} with the SS method, we follow the protocol used in \cite{zhang2018variational}. 

As we are benchmarking sophisticated versions of the phylogenetic inference CSMC \cite{wang2015bayesian}, it seems relevant to include a vanilla CSMC. We define the vanilla CSMC as one that proposes branch lengths using the model prior and merges subtrees uniformly. There are no learnable parameters in the vanilla CSMC. 

Indeed, there exist particle MCMC based approaches for parameterizing the CSMC, e.g., \cite{wang2015bayesian} or  \cite{wang2021particle}; however, these have not previously been benchmarked on the datasets used here, and the latter uses the K2P substitution model. As such, we leave these comparisons for future work and constrain ourselves to VI based CSMCs.

Following \cite{moretti2021variational}, we give all CSMC methods $K=2048$ particles. The parameters of VCSMC, VaiPhy, and $\phi$-CSMC are selected based on their best iteration, measured in LL. Using these parameters, we compute the LL estimates averaged over ten random seeds. We run VaiPhy and $\phi$-CSMC for $200$ iterations and evaluate them on Eq. \eqref{eq:iwelbo} and Eq. \eqref{eq:logZ}, respectively. Vanilla CSMC is also evaluated using Eq. \eqref{eq:logZ}.

In Table \ref{tab:ds_results}, we provide the mean LL scores and standard deviations. On all datasets except DS2, our $\phi$-CSMC is the superior CSMC method (highlighted in red). It also produces significantly smaller variations in its LL estimates compared to VCSMC and the Vanilla CSMC. On DS2, the Vanilla CSMC is surprisingly the winning CSMC.

Overall, MrBayes SS is the superior method on these datasets, followed by VBPI-NF. VaiPhy, despite its heavily penalizing prior distribution over tree topologies (see Sec. \ref{sec:vaiphy2csmc}), still outperforms VCSMC and Vanilla CSMC on all datasets but DS2, with extremely low-variance estimates of the LL.

\begin{table*}[t]
\caption{Real dataset LL estimates for the different methods. Our $\phi$-CSMC is, on average, the best-performing CSMC method (\textcolor{red}{red}). Overall, MrBayes SS achieves the highest LL estimates (\textbf{bold}). *Not available due to unresolved memory issues.}
\label{tab:ds_results}
\vskip 0.15in
\begin{center}
\resizebox{\textwidth}{!}{
\begin{small}
\begin{sc}
\begin{tabular}{lccrrrrr}
\toprule
Data Set & (\#Taxa, \#Sites) & MrBayes SS & VBPI-NF & VCSMC (JC) & Vanilla CSMC & VaiPhy & $\phi$-CSMC \\
\midrule
DS1 & (27, 1949) & \textbf{-7032.45 $\pm$ 0.15} & -7108.40 $\pm$ 0.11 & -9180.34 $\pm$ 170.27 & -8306.76 $\pm$ 166.27 & -7490.54 $\pm$ 0 & \textcolor{red}{-7290.36 $\pm$ 7.23} \\
DS2 & (29, 2520) & \textbf{-26363.85 $\pm$ 0.33} & -26367.70 $\pm$ 0.03 & -28700.7 $\pm$ 4892.67 & \textcolor{red}{-27884.37 $\pm$ 226.60} & -31203.44 $\pm$ 3E$^{-12}$ & -30568.49 $\pm$ 31.34 \\
DS3 & (36, 1812) & \textbf{-33729.60 $\pm$ 0.74} & -33735.09 $\pm$ 0.05 & -37211.20 $\pm$ 397.97 & -35381.01 $\pm$ 218.18 & -33911.13 $\pm$ 7E$^{-12}$ & \textcolor{red}{-33798.06 $\pm$ 6.62} \\
DS4 & (41, 1137) & \textbf{-13292.37 $\pm$ 1.42} & -13329.93 $\pm$ 0.11 & -17106.10 $\pm$ 362.74 & -15019.21 $\pm$ 100.61 & -13700.86 $\pm$ 0 & \textcolor{red}{-13582.24 $\pm$ 35.08} \\
DS5 & (50, 378) & \textbf{-8192.96 $\pm$ 0.71} & -8214.61 $\pm$ 0.40 & -9449.65 $\pm$ 2578.58 & -8940.62 $\pm$ 46.44 & -8464.77 $\pm$ 0 & \textcolor{red}{-8367.51 $\pm$ 8.87} \\
DS6 & (50, 1133)& \textbf{-6571.02 $\pm$ 0.81} & -6724.36 $\pm$ 0.37 & -9296.66 $\pm$ 2046.70 & -8029.51 $\pm$ 83.67 & -7157.84 $\pm$ 0 & \textcolor{red}{-7013.83 $\pm$ 16.99} \\
DS7 & (64, 1008) & \textbf{-8438.78 $\pm$ 73.66} & -8650.49 $\pm$ 0.53 & N/A$^*$ & -11013.57 $\pm$ 113.49 & -9462.21 $\pm$ 1E$^{-12}$ & \textcolor{red}{-9209.18 $\pm$ 18.03} \\
\bottomrule
\end{tabular}
\end{sc}
\end{small}}
\end{center}
\vskip -0.1in
\end{table*}

\subsection{Wall-Clock Time and Memory Comparisons}
Being a mean-field VI algorithm, VaiPhy truly enjoys the related benefits in terms of speed, which we demonstrate in this section. The experiments are performed on a high-performance computing cluster node with two Intel Xeon Gold 6130 CPUs with 16 CPU cores each. Each node in the cluster has 96 GiB RAM. 

\begin{figure}[ht]
\begin{center}
\includegraphics[width=6.5cm]{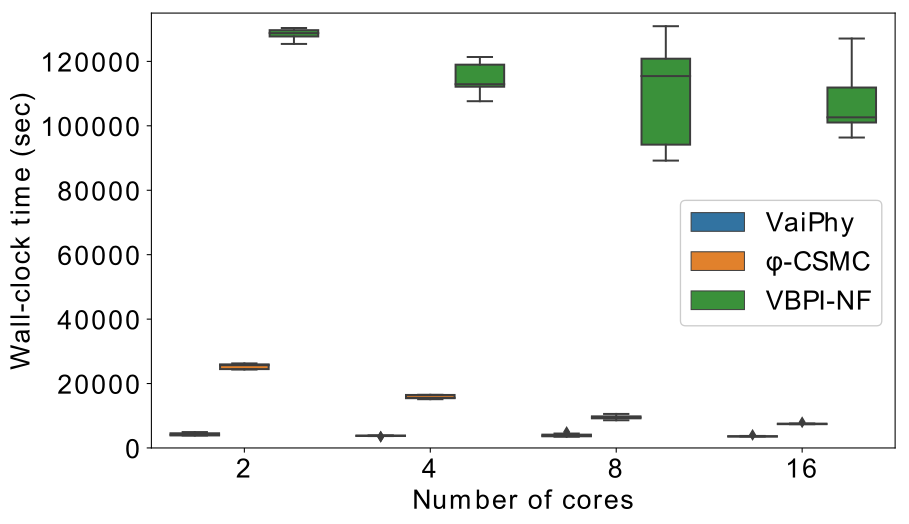} \hskip 0.2in
\includegraphics[width=6.5cm]{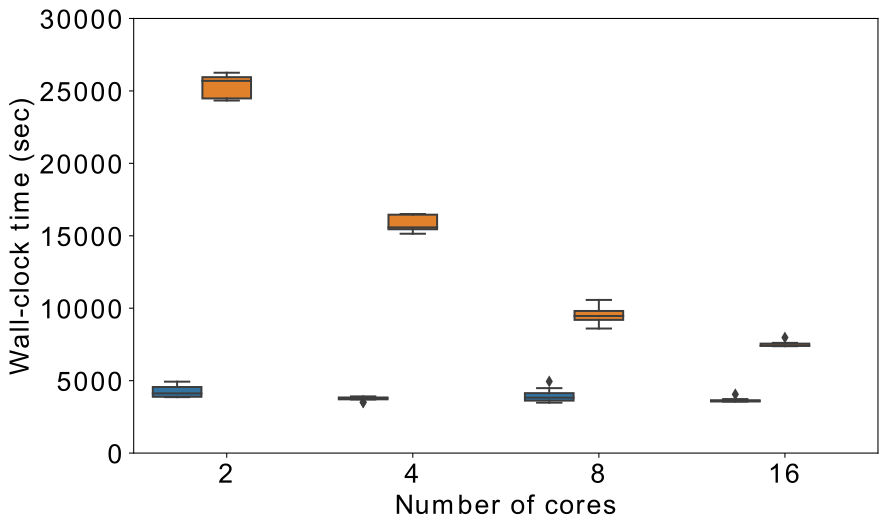}
\caption{Wall-clock runtime results on DS1 data set with $\{2, 4, 8, 16\}$ cores. Left) Runtime results for VaiPhy, $\phi$-CSMC, and VBPI-NF. Right) A zoomed-in version of the top figure showing runtime results for VaiPhy and $\phi$-CSMC. Clearly, VaiPhy is consistently faster than its benchmarks.}
\label{fig:runtime}
\end{center}
\vskip -0.2in
\end{figure}

We run VaiPhy for 200 iterations, and 128 trees are sampled using SLANTIS to compute the expectations in Eq. \eqref{eq:upd z} and \eqref{eq: upd B}. To compute $\mathcal{L}_L$, 3000 trees are sampled from SLANTIS after 200 iterations. The $\phi$-CSMC results include the aforementioned VaiPhy runs and 10 additional independent CSMC runs with 2048 particles. VBPI-NF runs for 400,000 iterations and uses 100,000 trees to estimate the lower bound after training.
Each experiment is repeated 10 times. The wall-clock time comparison of the methods on DS1 is presented in Fig. \ref{fig:runtime}.

We were not able to run VCSMC on less than 16 cores due to the method producing memory issues on DS1. However, for 16 cores, their runtime was around $6840$ seconds, with too small variations to show in a boxplot. VCSMC is thus faster than $\phi$-CSMC on 16 cores but still slower than VaiPhy. 

VaiPhy is remarkably faster than other variational methods for each number of cores investigated. Although VBPI-NF produces impressive LL estimates, it is considerably slower than VaiPhy and the CSMC methods (all CSMCs are not shown in the figure). Additionally, using two cores, the wall-clock time required by VBPI-NF to arrive at VaiPhy's approximation accuracy was close to 11x that of VaiPhy on DS1. For completeness, we compared wall-clock times with MrBayes, although this is not appropriate given that the software has been highly optimized over decades and is written in C. And, indeed, MrBayes is the fastest algorithm when allowed multiple cores. However, for a single core on DS1, MrBayes ran for approximately $100$ minutes while VaiPhy only needed an impressive $83$ minutes on average for the same setup. 

Finally, we discuss memory requirements. Our main variational baseline, VBPI-NF, used 8.4 million neural network weights and biases on DS1, while the number of learnable parameters (i.e., the dimensions of $\phi$ and $q(Z|X)$) in VaiPhy was 0.8 million; less than 10\% of the former. 

\section{Conclusion}
We have presented a modular framework for Bayesian phylogenetic inference, building on three novel methods: \textit{VaiPhy}, \textit{SLANTIS}, and the \textit{JC sampler}. At the heart of our framework lies $\phi$, the estimated number of mutations matrix. Although $\phi$ was learned in a multifurcating tree space with labeled internal vertices, we showed that the evolutionary distances also served useful in the bifurcating $X$-tree space. We demonstrated this by using $\phi$ to parameterize two new proposal distributions for the CSMC, also introduced by us in this work; we referred to this new algorithm as $\phi$-CSMC. Our $\phi$-CSMC proved to be the best-performing CSMC in terms of marginal log-likelihood estimation on real datasets. Additionally, VaiPhy was up to \textit{25 times faster} than its baselines on real data.

 Speed is essential for making Bayesian phylogenetics a reality for future large data sets, in particular, single-cell data sets in medical investigations. Moreover, speed and the opportunity to leverage the VI methodology for phylogeny, as provided by our framework, i.e., our graph-based representation of the VI distribution over trees, $\phi$, opens up novel research opportunities. For instance, recently, ensemble and mixture models in VI have received a lot of attention \cite{kviman2022multiple, morningstar2021automatic}. VaiPhy could obtain many ensemble components in parallel with low computational cost and thus has great potential to benefit from these findings. Moreover, more mixture components appear to imply significant improvements in terms of negative log-likelihood scores when learned by maximizing a multiple importance sampling-based objective, \textit{MISELBO} \cite{kviman2022learning}.
 This motivates researching how VaiPhy could be fitted using a MISELBO objective.

For many potential applications, models other than the JC69 model will be more suitable for the evolution that takes place over the edges of the phylogenetic tree. However, our JC sampler provides a great starting point. Reasonably, the JC sampler should serve as a better base distribution for normalizing flows than those currently employed \cite{Zhang0if}. Furthermore, in this work, the JC sampler samples the branch lengths independent of the tree topology. We expect that tree topology dependent branch sampling will enhance VaiPhy's performance. This will be investigated in our future work.

VI can facilitate Bayesian analysis of models that integrate phylogeny with other tasks that usually are performed as pre-processing or post-processing steps. The most important such example may very well be sequence alignment. Although Bayesian phylogenetic inference is currently considered state-of-the-art, this inference does not take the observed sequences associated with the taxa as input but a so-called multiple sequence alignment (MSA) of these, which is obtained by applying a heuristic to the sequences. This implies that the entire uncertainty related to the multitude of possible MSAs remains uncharacterized after a standard Bayesian phylogenetic inference, i.e., it does not affect the obtained posterior at all. For single-cell phylogenetic analysis of cancer cells, pre-processing may be even more involved. For instance, the so-called direct library preparation single-cell DNA data \cite{Zahn2017-nj} for tumors require a pre-processing step in which the cells are clustered into cancer clones. For these clones, it is then desirable to perform a phylogenetic inference. Interestingly, the VI has already been launched for Bayesian clustering for this pre-processing step \cite{safinianaini2020copymix}.

In conclusion, we have presented a VI algorithm for Bayesian phylogeny that, together with its building blocks, unlocks the potential of VI for fast Bayesian analysis in phylogeny and integrated phylogenetic analysis. The by now 25 year old methodology of MCMC based Bayesian phylogenetic analysis is certainly a mature and optimized methodology with great accuracy. However, VaiPhy is faster than MrBayes on a single core, and the underlying methodology is bound to lead to a sequence of increasingly faster and more accurate VI based phylogenetic software.  Moreover, it paves the way for integrated VI based phylogenetic inference.

\begin{ack}
First, we acknowledge the insightful comments provided by the reviewers, which have helped improve our work. This project was made possible through funding from the Swedish Foundation for Strategic Research grant BD15-0043, and from the Swedish Research Council grant 2018-05417\_VR. The computations and data handling were enabled by resources provided by the Swedish National Infrastructure for Computing (SNIC), partially funded by the Swedish Research Council through grant agreement no. 2018-05973.

The authors declare no competing interests.
\end{ack}

\bibliography{neurips_2022}
\bibliographystyle{abbrv}

\newpage
\section*{Checklist}


\begin{enumerate}

\item For all authors...
\begin{enumerate}
  \item Do the main claims made in the abstract and introduction accurately reflect the paper's contributions and scope?
    \answerYes{}
  \item Did you describe the limitations of your work?
    \answerYes{We discuss the extension to labelled multifurcating tree space and the requirement of the post-processing step that is required to make fair comparisons in bifurcating space.}
  \item Did you discuss any potential negative societal impacts of your work?
    \answerNA{}
  \item Have you read the ethics review guidelines and ensured that your paper conforms to them?
    \answerYes{}
\end{enumerate}

\item If you are including theoretical results...
\begin{enumerate}
  \item Did you state the full set of assumptions of all theoretical results?
    \answerYes{}
    \item Did you include complete proofs of all theoretical results?
    \answerYes{We show the conjugacy of JC sampler and we provide proof of natural gradient in the Appendix.}
\end{enumerate}

\item If you ran experiments...
\begin{enumerate}
  \item Did you include the code, data, and instructions needed to reproduce the main experimental results (either in the supplemental material or as a URL)?
    \answerYes{We will provide the codes in the supplementary materials.}
  \item Did you specify all the training details (e.g., data splits, hyperparameters, how they were chosen)?
    \answerYes{In the Experiments section we mention the choice of parameters. In Appendix, we provide code segments to show how we ran the benchmark methods.}
        \item Did you report error bars (e.g., with respect to the random seed after running experiments multiple times)?
    \answerYes{In Table 1 we include the standard deviations and in Figure 3 we include the error bars.}
        \item Did you include the total amount of compute and the type of resources used (e.g., type of GPUs, internal cluster, or cloud provider)?
    \answerYes{We wrote the hardware specifications in the runtime experiments; however, we didn't include the total amount of time required for all datasets.}
\end{enumerate}

\item If you are using existing assets (e.g., code, data, models) or curating/releasing new assets...
\begin{enumerate}
  \item If your work uses existing assets, did you cite the creators?
    \answerYes{We cited all software and also provided GitHub links to the repositories in the Appendix.}
  \item Did you mention the license of the assets?
    \answerNA{}
  \item Did you include any new assets either in the supplemental material or as a URL?
    \answerYes{The code is available in the supplementary material.}
  \item Did you discuss whether and how consent was obtained from people whose data you're using/curating?
    \answerNA{}
  \item Did you discuss whether the data you are using/curating contains personally identifiable information or offensive content?
    \answerNA{}
\end{enumerate}

\item If you used crowdsourcing or conducted research with human subjects...
\begin{enumerate}
  \item Did you include the full text of instructions given to participants and screenshots, if applicable?
    \answerNA{}
  \item Did you describe any potential participant risks, with links to Institutional Review Board (IRB) approvals, if applicable?
    \answerNA{}
  \item Did you include the estimated hourly wage paid to participants and the total amount spent on participant compensation?
    \answerNA{}
\end{enumerate}

\end{enumerate}

\newpage

\begin{center}
\rule{\textwidth}{4pt} \\ \vspace{0.25in} 
\fontsize{17}{1}\textbf{VaiPhy: a Variational Inference Based Algorithm for Phylogeny \\ Appendix} \\
\vspace{0.25in} \rule{\textwidth}{1pt}
\end{center}
\setcounter{page}{1}

\appendix
\section{The VaiPhy Algorithm}
\label{appendix:vaiphy_general}

\subsection{Update Equation Details}
\label{appendix:vaiphy}

The update equations of VaiPhy follow the standard mean-field VI updates. The variational distribution factorizes over the tree topology, ancestral sequences of each latent vertex, and the branch lengths of all edges in the graph (Eq. \eqref{eq:q_dist}); 

\begin{align}
\label{eq:q_dist}
    \qphi(\tau, \mathcal{B}, Z|X) &= \qphi(\mathcal{B}|X) \qphi(\tau|X) \qphi(Z|X) \\
    &= \prod_{e\in \mathcal{G}} \qphi(b(e)|X) \qphi(\tau|X) \prod_{i\in I(\mathrm{A})} q(Z_i|X).  \\
\end{align}

For brevity, we denote the set of neighbors of node $i$ with $N_{\tau}(i)$ and their sequences with $Y_{N_{\tau}(i)}$ in this section. Furthermore, $\neg i$ is the set of nodes except node $i$, and $C$ is a constant. 

\textbf{Ancestral sequence update equation}

\begin{align}
    \log q^*(Z_i|X) &\propto \mathbb{E}_{\qphi(\tau, \mathcal{B}, Z_{\neg i}|X)} \left[ \log p_\theta(X, Z, \mathcal{B}, \tau) \right]\\
    &= \sum_{\tau} \int_{\mathcal{B}} \sum_{Z_{\neg i}} \qphi(\tau|X) \qphi(\mathcal{B}|X) q(Z_{\neg i}|X) \left[ \log p_\theta(X, Z_i, Z_{\neg i}, \mathcal{B}, \tau) \right] d\mathcal{B} \\ 
    &= \sum_{\tau} \int_{\mathcal{B}} \qphi(\tau|X) \qphi(\mathcal{B}|X) \left[ \sum_{Y_j \in Y_{N_\tau(i)}} q(Y_j|X) \log p_\theta(Y_j | Z_i, b(i,j), \tau) \right] d\mathcal{B} + C \\ 
    &\equptoadd \mathbb{E}_{\qphi(\tau, \mathcal{B}| X)} \left[ \sum_{Y_j \in Y_{N_\tau(i)}} q(Y_j|X) \log p_\theta(Y_j | Z_i, b(i,j), \tau)  \right] \\ 
\end{align}

If $j$ is an observed node, $q(Y_j | X) = {\mathbbm{1}\{Y_j= X_j\}}$. 

\textbf{Tree topology update equation}
\begin{align}
    \log q^*(\tau|X) &\propto \mathbb{E}_{\qphi(\mathcal{B}, Z|X)}\left[
    \log p_\theta(X, Z, \mathcal{B}, \tau)
    \right]\\
    &= \sum_{Z} \int_{\mathcal{B}} \qphi(\mathcal{B}|X) q(Z|X) \log p_\theta(X, Z, \mathcal{B}, \tau) d\mathcal{B} \\
    &= \sum_{Z} \int_{\mathcal{B}} \qphi(\mathcal{B}|X) q(Z|X) \log p_\theta(X, Z | \mathcal{B}, \tau) d\mathcal{B} + \log p_\theta(\tau) + C \\
    &= \sum_{(i,j) \in E(\tau)} \sum_{Z_i, Z_j} \int_{b(i,j)} \qphi(b(i,j)|X) q(Z_i, Z_j|X) \log p_\theta(Z_i | Z_j, b(i,j), \tau) db(i,j) \\
    &\quad\quad+ \log p_\theta(\tau) + C \\
    &\equptoadd \mathbb{E}_{\qphi(\mathcal{B}|X)} \left[
    \sum_{(i,j) \in E(\tau)} \sum_{Z_i, Z_j} q(Z_i, Z_j|X) \log p_\theta(Z_i | Z_j, b(i,j), \tau)
    \right] \\
\end{align}

\textbf{Branch length update equation} 
\begin{align}
    \log \; &q^*(b(i,j)|X) \\
    &\propto \mathbb{E}_{\qphi(\tau, \mathcal{B}_{\neg e(i,j)}, Z|X)} \left[ \log p_\theta(X, Z, \mathcal{B}, \tau) \right]\\
    &= \sum_{\tau} \sum_{Z} \int_{\mathcal{B}_{\neg e(i,j)}} \qphi(\tau|X) \qphi(Z|X) \qphi(\mathcal{B}_{\neg e(i,j)}|X) \left[ \log p_\theta(X, Z, b(i,j), \mathcal{B}_{\neg e(i,j)}, \tau) \right] d\mathcal{B}_{\neg e} \\
    &= \sum_{\tau} \qphi(\tau|X) \sum_{Z_i, Z_j}  \qphi(Z_i, Z_j|X) \left[ \log p_\theta(Z_i | Z_j, b(i,j), \tau) \right] + \log p_\theta(b(i,j)) + C \\
    &\equptoadd \mathbb{E}_{\qphi(\tau|X)} \left[ \sum_{Z_i, Z_j}  \qphi(Z_i, Z_j|X) \log p_\theta(Z_i | Z_j, b(i,j), \tau) \right] + \log p_\theta(b(i,j)) \\
\end{align}

In the experiments, we observed that using the branch length that maximizes the tree likelihood during optimization provided better results. Hence, during the training of VaiPhy, we used a maximum likelihood heuristic to update the branch lengths given a tree topology. After the training, we used the tree topologies sampled from SLANTIS and corresponding branch lengths sampled from the JC sampler to compute IWELBO.

\subsection{Neighbor-Joining Initialization}
\label{appendix:nj_init}

We utilize the NJ algorithm to initialize VaiPhy with a reasonable state. The sequence data is fed into BIONJ, an NJ algorithm, to create an initial reference phylogenetic tree using the PhyML software, version 3.3.20200621 \cite{gascuel1997bionj,guindon2010new}. The branch lengths of the NJ tree are optimized with the same software. An example script to run PhyML is shown below. 

\begin{verbatim}
phyml -i DS1.phylip -m JC69 --r_seed vbpi_seed -o l -c 1
\end{verbatim}


The marginal likelihoods of internal vertices, $p_\theta(Z_i|X, \mathcal{B}, \tau) \; \forall i \in I(A)$, are used to initialize the latent ancestral sequences, $\qphi(Z_i | X)$. The optimized branch lengths are used as the initial set of lengths for $e \in E(\tau)$. The lengths of the edges that are not present in the NJ tree are initialized by computing the shortest path between the vertices using the Floyd-Warshall algorithm \cite{floyd1962algorithm}.

\subsection{Graphical Model}
\label{appendix:graphical model}
When training VaiPhy, we used the following branch length prior
\begin{equation}
    p_\theta(\mathcal{B}|\tau) = \prod_{e\in E(\tau)}p_\theta(b(e)),
\end{equation}
where $p_\theta(b(e))=\text{Exp}(10)$, and a uniform prior over tree topologies (in $\mathrm{A}$):
\begin{equation}
    p_\theta(\tau) = \frac{1}{N ^{N - 2}}.
\end{equation}
The prior when considering unrooted, bifurcating $X$-trees (which is the case for VBPI-NF):
\begin{equation}
    p'(\tau) = \frac{1}{(2|X| - 3)!!},
\end{equation}
and for rooted, bifurcating $X$-trees
\begin{equation}
    p''(\tau) = \frac{1}{(2|X| - 5)!!}.
\end{equation}

\newpage
\section{SLANTIS}
\label{appendix:slantis}

Here we provide two algorithmic descriptions of SLANTIS. Note that in Eq. \eqref{eq:weight matrix}, the weight matrix $W$ is in logarithmic scale. However, in Alg. \ref{alg:slantis} and \ref{alg:slantis-propagation}, $W$ is in normal scale. 

\begin{algorithm}[H]
\caption{SLANTIS in pseudocode}
\label{alg:slantis}
\begin{algorithmic}[1]
\STATE {\bfseries Input:} $\phi, W$
\STATE $\mathcal{G} \leftarrow$ graph spanned by $W$
\STATE $\mathcal{G_I} \leftarrow \mathcal{G} \setminus \Lambda$
\STATE Initialize $T_1$, e.g. by MST($W(\mathcal{G_I})$)
\STATE Initialize $\tau \leftarrow T_1$, $s(\tau) \leftarrow 1$, $\mathcal{M} \leftarrow \{ \}$ and $s \leftarrow 1$
\WHILE{$|\mathcal{M}| < |V(\mathcal{G_I})| - 1$}
    \STATE $\mathcal{R} \leftarrow \mathcal{G_I} \setminus T_s$
    \STATE $W_1 \leftarrow W(\mathcal{R})$
    \STATE $T_{s+1} \leftarrow$ MST($W_1$)
    \STATE pass $T_s, T_{s+1}, \tau, W, \mathcal{M}$ to Algorithm \ref{alg:slantis-propagation}
    \STATE $\tau, \mathcal{M}$ updated in Alg. \ref{alg:slantis-propagation}
    \STATE $s(\tau) \leftarrow s(\tau) \times r$, from return value $r$ of Alg. \ref{alg:slantis-propagation}
    \STATE $T_s \leftarrow T_{s+1}$, $s \leftarrow s + 1$
\ENDWHILE
\STATE Sample leaf connections
\FOR{$u \in V(\Lambda)$}
    \STATE $i \leftarrow \text{indexes of internal vertices in } \mathcal{G}$ 
    \STATE $p_1, ..., p_n \leftarrow \frac{W(u, i_1)}{\Sigma W(u, i)} , ..., \frac{W(u, i_n)}{\Sigma W(u, i)}$
    \STATE $v \leftarrow$ Categorical($p_1, ..., p_n$)
    \STATE add $(u,v)$ to $\tau$ 
    \STATE $s(\tau) \leftarrow s(\tau) \times p_v$
\ENDFOR
\STATE \textbf{return} $\tau$, $s(\tau)$
\end{algorithmic} 
\end{algorithm}

\newpage
\begin{algorithm}[H]
\caption{SLANTIS - propagation}
\label{alg:slantis-propagation}
\begin{algorithmic}[1]
\STATE {\bfseries Input:} $t_0, t_1, \tau, W, \mathcal{M}$
   \STATE Initialize $r \leftarrow 1$
   \STATE $I_1 \leftarrow sort(t_1)$, e.g. by sorting $W(t_1)$ in descending order and selecting the indices.
   \FOR{$e \in t_0$}
       \IF{$e \in \tau$}
       \STATE Set $\tau_0, \tau_1 \text{ such that } e = cut(\tau_0, \tau_1)$ 
       \FOR{$e_1 \in I_1$}
            \IF{$e_1 \text{ connects } \tau_0, \tau_1$}
                \STATE $e^{*} \leftarrow e_1$
                \STATE \textbf{break} inner for loop
            \ENDIF
        \ENDFOR
        \STATE $r_e = \frac{W(e)}{W(e) + W(e^*)}$
        \ELSIF{$e \notin \tau$}
            \STATE add $e$ to $\tau$ creating new graph $\tau^*$ with cycle $\mathcal{C}$
            \STATE $c \leftarrow \argmin(W(\mathcal{C} \setminus \{\mathcal{M} \cup e^*\}))$
            \IF {c = \{\} }
                \STATE \textbf{continue}
            \ENDIF
            \STATE remove $c$ from $\tau^*$
            \STATE $r_e = \frac{W(e)}{W(e) + W(c)}$
        \ENDIF
        
        \STATE $u \leftarrow \text{Uniform}(0,1)$
        \IF{$e \in \tau$ \textbf{and} $r_e < u$}
        \STATE $\tau \leftarrow \tau_0 \cup e^* \cup \tau_1$
        \ELSIF{$e \in \tau$ \textbf{and} $r_e \geq u$}
        \STATE $M \leftarrow M \cup e$
        \STATE $r \leftarrow r \times r_e$
        \ELSIF{$e \not \in \tau$ \textbf{and} $r_e < u$}
        \STATE \textbf{continue}
        \ELSIF{$e \not \in \tau$ \textbf{and} $r_e \geq u$} 
        \STATE $\tau \leftarrow \tau^*$
        \STATE $M \leftarrow M \cup e$
        \STATE $r \leftarrow r \times r_e$
        \ENDIF
        \IF{$|\mathcal{M}| = |V(\mathcal{G_I})| - 1$}
            \STATE \textbf{break} for loop
        \ENDIF
    \ENDFOR
    \STATE \textbf{return} $\tau$, $\mathcal{M}, r$
\end{algorithmic} 
\end{algorithm}

\begin{figure}
    \centering
    \includegraphics{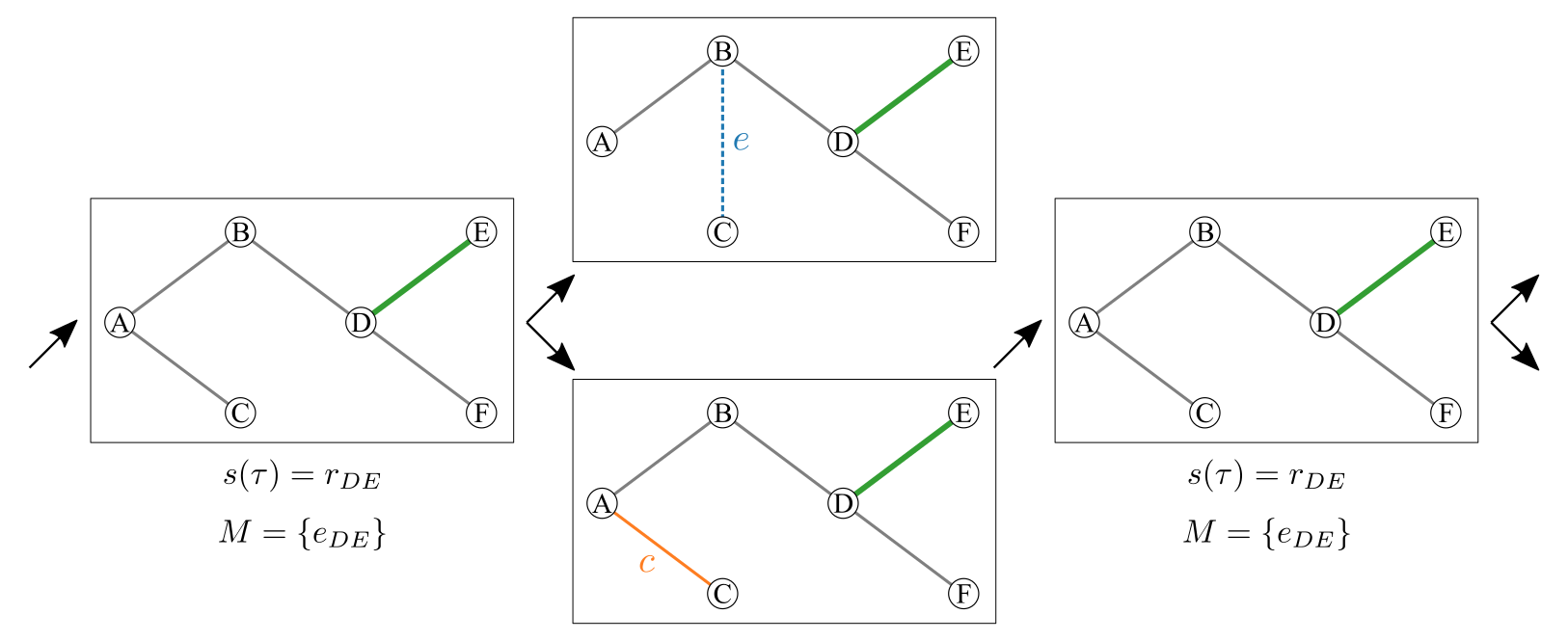}
    \caption{Propagation step of SLANTIS, a case where $e \not \in \tau$ and $c \in \tau$. The Bernoulli r.v. is $r_e = W(e) / (W(e) + W(c))$. $e$ is rejected, therefore $\tau$, $s(\tau)$ and $M$ remain unchanged.}
    \label{fig:slantis_prop_1}
\end{figure}

\begin{figure}
    \centering
    \includegraphics{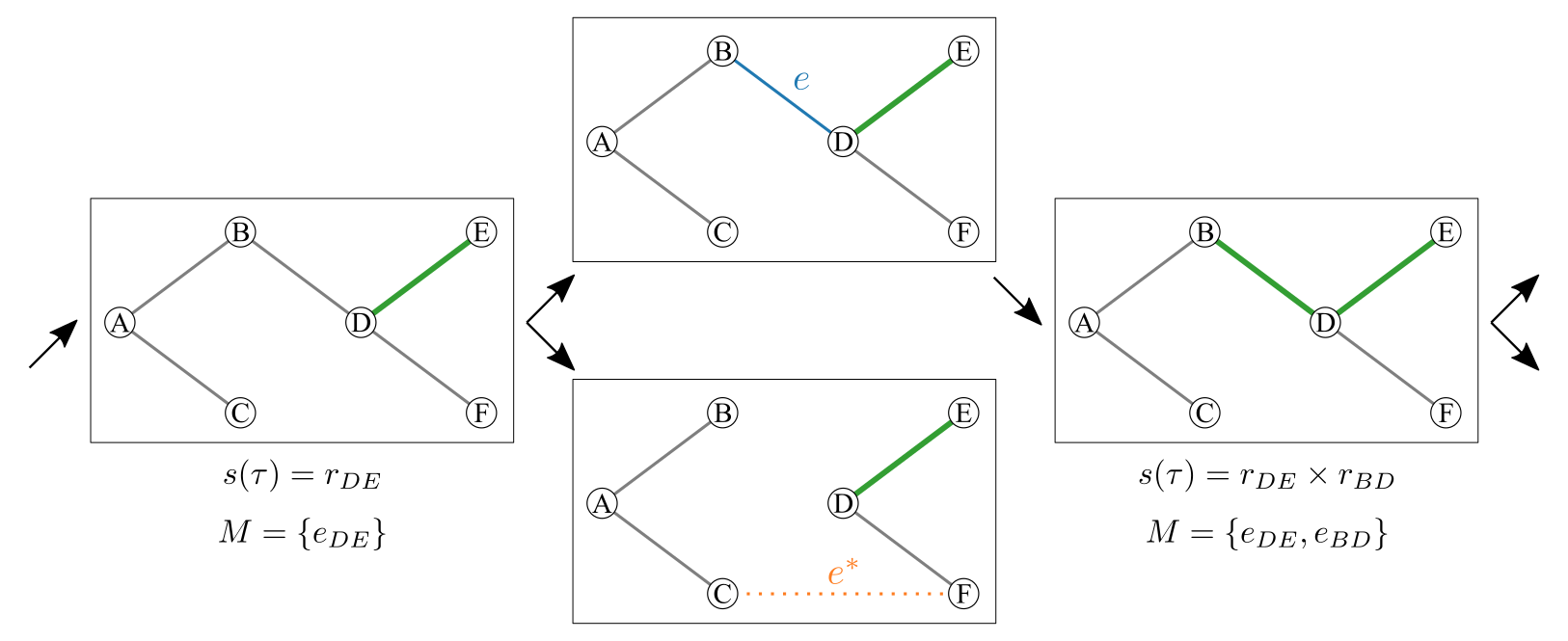}
    \caption{Propagation step of SLANTIS, a case where $e \in \tau$ and $e^* \not \in \tau$. The Bernoulli r.v. is $r_e = W(e) / (W(e) + W(e^*))$. $e$ is accepted, therefore $\tau$, $s(\tau)$ and $M$ are updated.}
    \label{fig:slantis_prop_2}
\end{figure}

\begin{figure}
    \centering
    \includegraphics{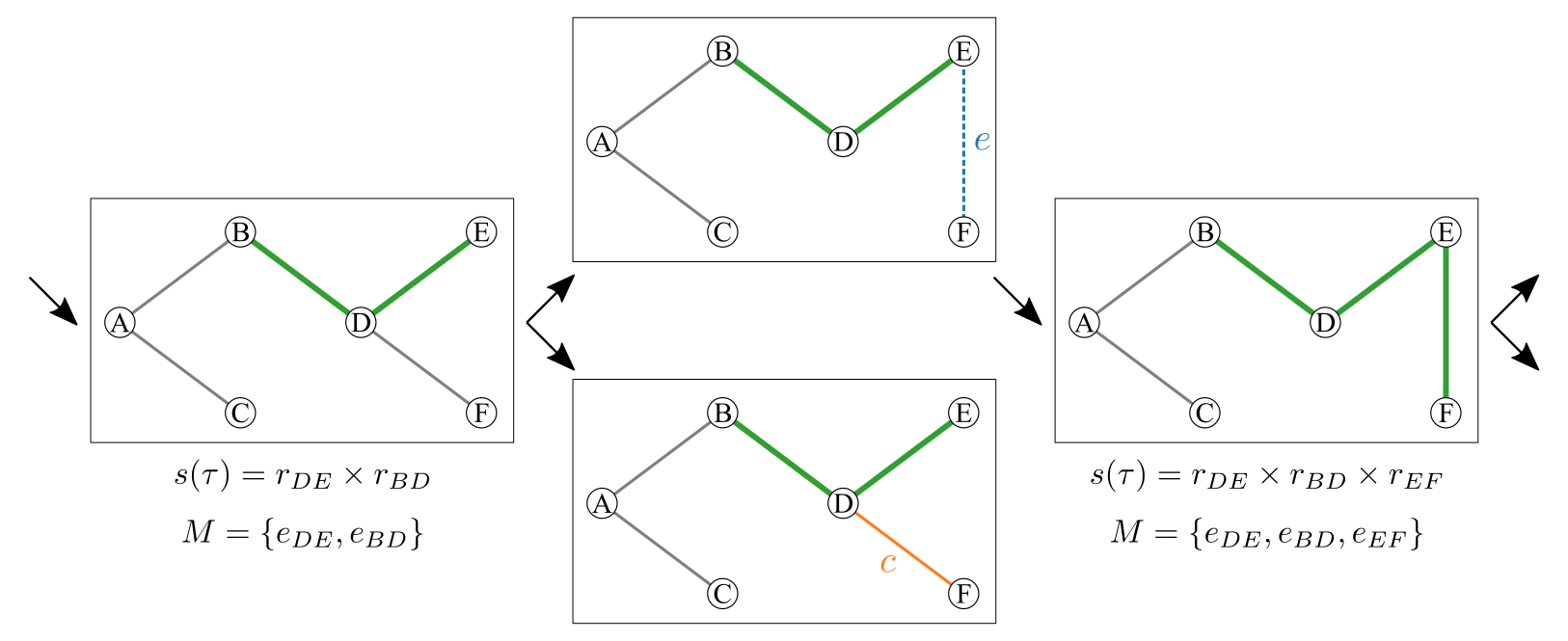}
    \caption{Propagation step of SLANTIS, a case where $e \not \in \tau$ and $c \in \tau$. The Bernoulli r.v. is $r_e = W(e) / (W(e) + W(c))$. $e$ is accepted, therefore $\tau$, $s(\tau)$ and $M$ are updated.}
    \label{fig:slantis_prop_3}
\end{figure}

\begin{figure}
    \centering
    \includegraphics{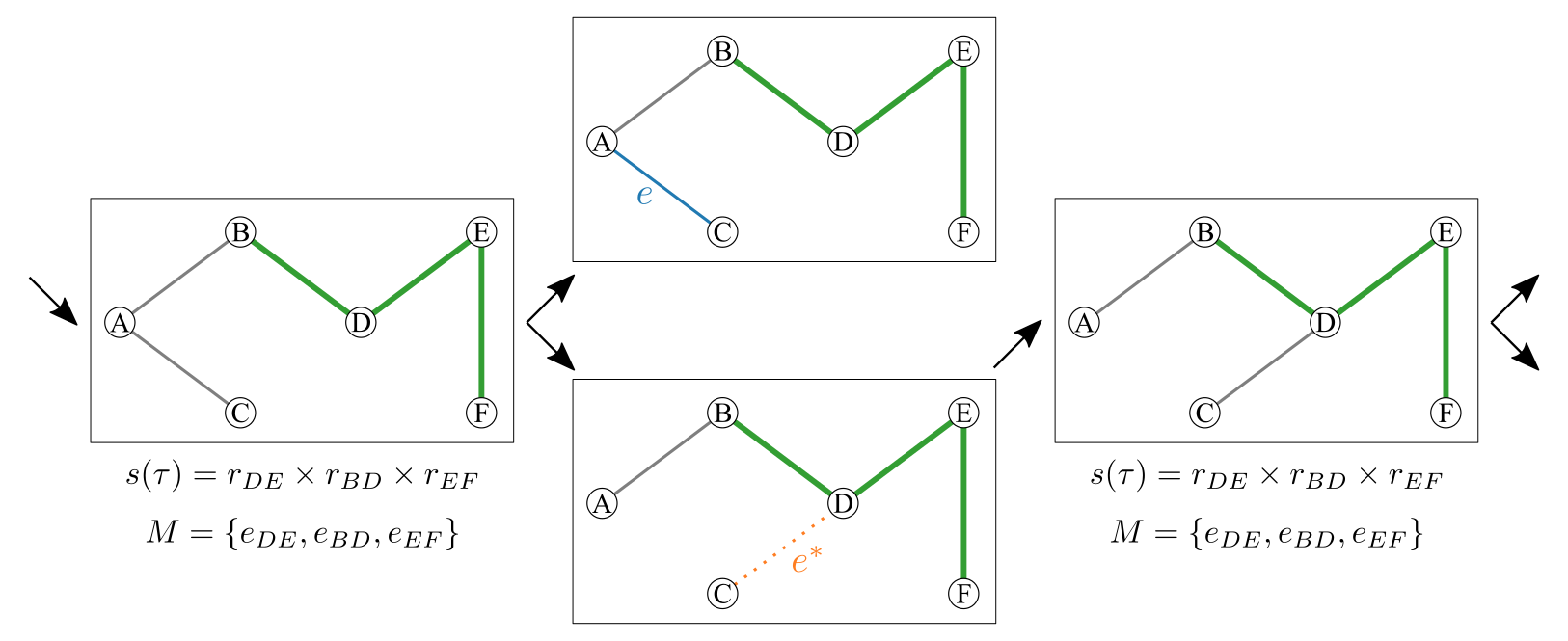}
    \caption{Propagation step of SLANTIS, a case where $e \in \tau$ and $e^* \not \in \tau$. The Bernoulli r.v. is $r_e = W(e) / (W(e) + W(e^*))$. $e$ is rejected, therefore $\tau$, $s(\tau)$ and $M$ remain unchanged.}
    \label{fig:slantis_prop_4}
\end{figure}

Fig. \ref{fig:slantis_prop_1}-\ref{fig:slantis_prop_4} show four different cases in SLANTIS propagation. In all of the figures, the left column is the current state of $\tau$, the middle column is two trees that are compared, and the right column is the selected tree. Solid lines indicate the edges in $\tau$, and bold \textbf{\textcolor{tab:green}{green}} lines are accepted edges (edges in $M$). The \textcolor{tab:blue}{blue} color indicates the edge we consider (\textcolor{tab:blue}{$e$}) 
and \textcolor{tab:orange}{orange} color indicates the alternative edge; either
\textcolor{tab:orange}{$c$} or \textcolor{tab:orange}{$e^*$}. Depending on the configuration, the Bernoulli random variable is either $r_e = W(e) / (W(e) + W(c))$ or $r_e = W(e) / (W(e) + W(e^*))$. In order to add an edge to $M$, the edge must have \textcolor{tab:blue}{blue} color (\textcolor{tab:blue}{$e$} $\in \tau$) and must get accepted. At the end of the SLANTIS algorithm, the spanning tree with bold \textbf{\textcolor{tab:green}{green}} edges, along with its sampled leaf connections, is returned. 

\newpage
\section{The JC Sampler}
\label{appendix:jc_sampler} 
In the JC Sampler, the probability of no substitutions is sampled from the Beta posterior distribution as follows;
$$p \sim \text{Beta}(p| M-\phi_{ij} + 1, \phi_{ij} + 1).$$

The Beta posterior distribution is parameterized by the Beta prior ($Beta(p|1,1)$ in this case), the sequence length ($M$), and the expected number of substitutions between two nodes ($\phi_{ij}$).  

Since the support of the Beta distribution is in $[0, 1]$, it is possible to sample $p<0.25$. However, according to the JC69 model, such a case corresponds to saturation, and the logarithm in Eq. \eqref{eq:f} becomes negative.

In practice, on the rare occasions where we encounter $p<0.25$, we discard that sample and draw another $p$ from the Beta posterior distribution. Yet, the user can adjust the prior distribution parameters to change the behavior of the posterior distribution (e.g., skew the distribution away from $p<0.25$). 

\section{CMSC Proposal Distributions}
\label{appendix:CSMC}
Here we go into detail about our novel CSMC proposals.
\subsection{The Branch Length Proposal}
\label{appendix:CSMC B proposal}
Here we outline our new proposal distribution for sampling sets of branch lengths in the CSMC,
$r_\phi(\mathcal{B}_{\rho}|\mathcal{B}_{\rho-1}, \mathcal{T}_{\rho})$. In practice, sampling the $\mathcal{B}_\rho$ means that, given the two edges most recently added  to $\mathcal{T}_\rho$ ($e$ and $e'$), we sample branch lengths using the JC sampler, $s_\phi(b(e))$, and concatenate $\{b(e), b(e')\}$ to the previous set, $\mathcal{B}_\rho = \{b(e), b(e')\} \cup \mathcal{B}_{\rho-1}$. The complicated part is identifying which entries in $\phi$ the edges correspond to.

We let the proposal distribution of branch lengths be a mixture of JC samplers while additionally allowing edge indices to be random variables:
\begin{equation}
    r_\phi(b, b', e, e') =  s_\phi(b|e) \; r(e|e') \; s_\phi(b'|e') \; r(e').
\end{equation}
Branch length $b$ given $e$ is sampled independently of $e'$, and vice versa; however, we assume that the edge indices are not mutually independent. If $e'$ is sampled first from the mixture, then $e$ is constrained to sharing the same parent node as $e$. We get the following priors
\begin{equation}
    r(e') = \sum_{i,j\in \mathcal{C'}(e')}\frac{c(i,j)}{C}\delta_{i,j}(u, v),
\end{equation}
where $c(i,j)$ is the number of occurrences of the edge in the pre-sampled SLANTIS trees, $C'$ is a normalizing constant, and $\delta_{i,j}(\cdot)$ is the Dirac delta with the mass on $i,j$. $\mathcal{C}(e')$ is the set of possible labeled edge indices for $e'$. E.g., if the root in $e'$ is a leaf node, then we only need to consider edges involving that labeled leaf. To estimate the index of the unlabeled edge $e'$, we thus sample from $r(e')$. Similarly, for $r(e|e')$, the set $\mathcal{C}(e|e')$ is even smaller, as we have estimated the label of the parent. 

The above results in the following likelihood function for the set of branch lengths:
\begin{equation}
    r_\phi(\mathcal{B}_{\rho}|\mathcal{B}_{\rho-1}, \mathcal{T}_{\rho}) = 
    \prod_{\bar{e}\in E(\mathcal{T}_{\rho})} \sum_{i,j}^N s_\phi(b(\bar{e}))r_\phi(i,j|\bar{e}),
\end{equation}
where $\bar{e}$ is the estimated edge index.

\subsection{The Merger Proposal}
\label{appendix:CSMC T proposal}

At each rank $\rho\in[1,R]$, the CSMC proposes a forest, $\mathcal{T}_{\rho}$, by \textit{merging} the roots of two subtrees in $\mathcal{T}_{\rho-1}$. 

Let $r_{1},\ldots,r_{R}$ be the roots of a forest $\mathcal{T}_{\rho-1}$.\footnote{For brevity, we focus on a single particle.} There are ${R \choose 2}$ way to merge the roots. In the case of vanilla CSMC, the probability of merging roots $r_i$ and $r_j$ is uniform, $p(r_i, r_j) = 1 / {R \choose 2}$. 

For $\phi$-CSMC, we sample $S$ trees from SLANTIS and corresponding branch lengths from JC sampler as a preprocessing step.\footnote{For the experiments, we set S=3,000.} Let $X(r_i)$ and $X(r_j)$ be the set of observations in the subtrees with roots $r_i$ and $r_j$ respectively. We define a function $O(X(r_i), X(r_j), \tau)$, that checks each edge in $\tau$ and returns $0$ if there is an edge that separates the observations into two disjoint sets such that one set of observations is $X(r_i) \cup X(r_j)$. The probability of merging subtrees with roots $r_i$ and $r_j$ is:

\begin{align}
    p(r_i, r_j) &\propto \sum_{s=1}^S p_\theta(X | \tau^s, \mathcal{B}^s) \mathbbm{1}\{O(X(r_i), X(r_j), \tau^s)=0\}
\end{align}

\section{Biological Data Sets}
\label{appendix:real_datasets}

The details of the biological data sets are presented in Table~\ref{tab:biological_data_details}. The former and up-to-date TreeBASE Matrix IDs are displayed \cite{vos2010treebase2}.

\begin{table}[h]
  \caption{Biological data set details}
  \label{tab:biological_data_details}
  \centering
  \begin{tabular}{llllll}
    \toprule
    & \multicolumn{2}{c}{TreeBASE Matrix ID} \\
    \cmidrule(r){2-3}
    Name & Legacy & Up-to-date Legacy & Taxa & Sites & Reference \\
    \midrule
    DS1  & M336  & M2017 & 27 & 1949 & \cite{hedges1990tetrapod} \\
    DS2  & M501  & M2131 & 29 & 2520 & \cite{garey1996molecular}\\
    DS3  & M1510 & M127  & 36 & 1812 & \cite{yang2003comparison} \\
    DS4  & M1366 & M487  & 41 & 1137 & \cite{henk2003laboulbeniopsis} \\
    DS5  & M3475 & M2907 & 50 & 378 & \cite{lakner2008efficiency} \\
    DS6  & M1044 & M220  & 50 & 1133 & \cite{zhang2001molecular} \\
    DS7  & M755  & M2261 & 64 & 1008 & \cite{rossman2001molecular} \\
    \bottomrule
  \end{tabular}
\end{table}

\newpage
\section{MrBayes Experiment}
\label{appendix:mrbayes-experiment}
Following the experimental setup in \cite{Zhang0if}, we ran MrBayes version 3.2.7 \cite{ronquist2012mrbayes} (written in C) with the stepping-stone algorithm and summarized the results of 10 independent runs. We set the number of generations to 10,000,000 and the number of chains per run to 4. The mean and the standard deviation of the marginal log-likelihood estimates are calculated by using the results reported in MrBayes output file. An example script to run MrBayes is shown below. 

\begin{verbatim}
BEGIN MRBAYES; 
set autoclose=yes nowarn=yes Seed=123 Swapseed=123;
lset nst=1; 
prset statefreqpr=fixed(equal); 
ss ngen = 10000000 nruns=10 nchains=4 printfreq=1000 \
samplefreq=100 savebrlens=yes filename=mrbayes_ss_out.txt; 
END;
\end{verbatim}


\section{VBPI-NF Experiment}
\label{appendix:vbpinf-experiment}
We used the codes in \hyperlink{https://github.com/zcrabbit/vbpi-nf}{the author's GitHub repository}, which is based on PyTorch, to reproduce their results on the biological data sets. VBPI-NF requires precomputed bootstrap trees to gather the support of conditional probability tables. For DS[1-4], we used the bootstrap trees provided by the authors in their code repository. For DS[5-7], following the guidelines in \cite{Zhang0if}, UFBoot is used to create 10 replicates of 10,000 bootstrap trees \cite{hoang2017ufboot}. An example script is shown below.

\begin{verbatim}
iqtree -s DS5 -bb 10000 -wbt -m JC69 -redo
\end{verbatim}


We chose RealNVP(10) variant for comparison. To reproduce the results on biological datasets, we used RealNVP flow type with 10 layers. The step size for branch length parameters is set to 0.0001.\footnote{We consulted the authors.} We introduced a seed parameter in their code to be able to reproduce the runs. We used default settings for the rest of the parameters. An example script is shown below. 

\begin{verbatim}
python main.py --dataset DS5 --flow_type realnvp --Lnf 10 \
--stepszBranch 0.0001 --vbpi_seed 1
\end{verbatim}


\section{VCSMC Experiment}
\label{appendix:VCSMC-experiment}
We attempted to reproduce the JC model results of VCSMC \cite{moretti2021variational} using \hyperlink{https://github.com/amoretti86/phylo}{the authors' GitHub repository} (based on TensorFlow 1) and default hyperparameters: 100 epochs, batch size 256, learning rate 0.001, branch length initialization $\ln (10)$. Similar to the VBPI-NF experiments, we added a seed argument for reproducibility. 

\begin{verbatim}
python runner.py --dataset hohna_data_1 --jcmodel True --num_epoch 100 \ 
--n_particles 2048 --batch_size 256 --learning_rate 0.001 --nested False --seed 1
\end{verbatim}

After running 100 epochs, the parameters $\lambda_{max}$ and $Q_{max}$ (although $Q_{max}$ is fixed for the JC model) of the epoch with maximum $\log \hat{p}(X)$ were selected. Then CSMC was run ten times using $\lambda_{max}$ and $ Q_{max}$. The results in Table \ref{tab:ds_results} correspond to the means and standard deviations for $\log \hat{p}(X)$ and $p(X|\tau,B)$ of these ten evaluation runs. 
In the VCSMC paper, the results are obtained by alternating optimization (updating $\lambda$ and $Q$) and inference (estimating $p(X)$ via the approximated ELBO, $\hat{\mathcal{L}_i}$) in 100 iterations,
ultimately reporting $\hat{\mathcal{L}}^\star = \max_i \{\hat{\mathcal{L}_i}\}_{i=1}^{100}$. Unfortunately, $\hat{\mathcal{L}}^\star$ will have a bias proportional to the variance of the selected $\hat{\mathcal{L}_i}$. 
Hence, $\mathbb{E}[\hat{\mathcal{L}}^\star]$ is not necessarily an ELBO. In practice, VCSMC produced $\hat{\mathcal{L}_i}$ with high variance (see Table \ref{tab:ds_results}), making it unclear if $\hat{\mathcal{L}}^\star$ results from a $\hat{\mathcal{L}_i}$ with a high mean or the variance. Instead, finding $\hat{\mathcal{L}}^\star$ and then rerunning the inference with the corresponding parameters for different seeds produces a fair benchmark; this is what we report.

The VNCSMC code in \hyperlink{https://github.com/amoretti86/phylo}{the authors' GitHub repository} gave rise to non-trivial memory issues that we couldn't resolve; hence, the VNCSMC method is excluded from this paper.

\newpage
\section{Proof of Convex Combination for Natural Gradient for Exponential Families}
\label{appendix:ng_proof}

The proof is two-folded: \textbf{Part 1}, where we prove that the Natural Gradient based VI update equation for Exponential Families (EF) is computed as a convex combination of old and new VI update equations; \textbf{Part 2}, where we prove that the VaiPhy distributions that should belong to EF, belong to EF.

\noindent \textbf{Part 1}: We basically prove what was stated in stochastic variational inference (SVI) \cite{hoffman2013stochastic} using SVI notation. Let $\mathcal{L}(q(\beta))$ denote the ELBO as a function of the variational distribution of the interest, as the goal is to maximize ELBO w.r.t. $q(\beta)$. We know that

\begin{align}
\mathcal{L}(q(\beta)) &= \mathbb{E}_q[\log p(\beta, x, Z, \alpha)] - \mathbb{E}_{q(\beta)}[\log q(\beta)] \\ &= \mathbb{E}_q[\log p(\beta | x, Z, \alpha) + \log p( x, Z, \alpha)] - \mathbb{E}_{q(\beta)}[\log q(\beta)]\\ & \equptoadd \mathbb{E}_q[\log p(\beta | x, Z, \alpha) ] - \mathbb{E}_{q(\beta)}[\log q(\beta)].
\end{align}

\noindent Note that for readability, \cite{hoffman2013stochastic} uses $q(\beta)$ instead of $q(\beta | \lambda)$, where $\lambda$ is the natural parameter. They assume that $q(\beta)$ and $p(\beta | x, Z, \alpha)$ belong to the same exponential family. However, we prove that they can merely belong to EF. Therefore, we have the following:

\begin{equation}
q(\beta) = H(\beta) e^{\lambda^T T(\beta) - A_g(\lambda)}, 
\end{equation}

\begin{equation}
p(\beta | x, Z, \alpha) = H(\beta) e^{\eta_g(x, Z, \alpha)^T T(\beta) - A_g(\eta_g(x, Z, \alpha))},
\end{equation}

\begin{equation}
\begin{aligned}
\mathcal{L}(q(\beta)) &= \mathbb{E}_{-q(\beta)}\big [ \mathbb{E}_{q(\beta)}[\log H(\beta) + \eta_g(x, Z, \alpha)^T T(\beta) - A_g(\eta_g(x, Z, \alpha))]\big ] \\
&\quad\quad- \mathbb{E}_{q(\beta)}[\log H(\beta)  + \lambda^T T(\beta) - A_g(\lambda)] \\
&\equptoadd \mathbb{E}_{-q(\beta)}\big [ \mathbb{E}_{q(\beta)}[\eta_g(x, Z, \alpha)^T T(\beta) -  A_g(\eta_g(x, Z, \alpha))]\big ] - \mathbb{E}_{q(\beta)}[ \lambda^T T(\beta) - A_g(\lambda)] \\
&\equptoadd \mathbb{E}_{-q(\beta)}\big [ \mathbb{E}_{q(\beta)}[\eta_g(x, Z, \alpha)^T T(\beta) ]\big ] - \mathbb{E}_{q(\beta)}[ \lambda^T T(\beta) - A_g(\lambda)] \\
& = \mathbb{E}_{-q(\beta)}\big [ \eta_g(x, Z, \alpha)^T \mathbb{E}_{q(\beta)}[ T(\beta) ]\big ] - \lambda^T \mathbb{E}_{q(\beta)}[ T(\beta)] + A_g(\lambda). 
\end{aligned}
\end{equation}

\noindent The final line above corresponds to Eq. 13 in \cite{hoffman2013stochastic}, except that they changed the measure so that they got $\mathbb{E}_{\lambda}[ T(\beta)]$ instead of $\mathbb{E}_{q(\beta)}[ T(\beta)]$; see the proof below.

\begin{lemma}

It holds that $\nabla_{\lambda} A_g(\lambda) = \mathbb{E}_{q(\beta)}[T(\beta)]$.

\end{lemma}

\textit{Proof.}

\noindent Since the derivative of the logarithm is the inverse of the variable, we have

\begin{align}
\nabla_{\lambda} A_g(\lambda) &= \nabla_{\lambda} \bigg\{ \log \int H(\beta) \exp\{\lambda^T T(\beta)\} d\beta \bigg \} = \frac{\nabla_{\lambda}  \int H(\beta) \exp\{\lambda^T T(\beta)\} d\beta }{\int H(\beta) \exp\{\lambda^T T(\beta)\} d\beta}.
\end{align}

\noindent Taking the derivative of the numerator in the above, we get

\begin{align}
\frac{\int H(\beta) T(\beta) \exp\{\lambda^T T(\beta)\} d\beta }{\int H(\beta) \exp\{\lambda^T T(\beta)\} d\beta}.
\end{align}

\noindent We now rewrite the above and then multiply and divide the expression by $\exp \{ A_g(\lambda)\}$, which results in

\begin{align}
\frac{\int H(\beta) T(\beta) \exp\{\lambda^T T(\beta)\} d\beta }{\int H(\beta) \exp\{\lambda^T T(\beta)\} d\beta} &= \int T(\beta) \frac{ H(\beta) \exp\{\lambda^T T(\beta)\} d\beta }{\int H(\beta) \exp\{\lambda^T T(\beta)\} d\beta} \\ & = \int T(\beta) \frac{H(\beta) \exp\{\lambda^T T(\beta)\} }{\exp\{A_g(\lambda)\}} \frac{\exp\{A_g(\lambda)\}}{\int H(\beta) \exp\{\lambda^T T(\beta)\} d\beta} d\beta.
\end{align}

\noindent We know that exponentiation of the log-normalizer gives us $\exp\{A_g(\lambda)\} = \int \exp \{ \lambda^T T(\beta)\} H(\beta) d \beta$. Therefore, we have

\begin{align}
&\int T(\beta) \frac{H(\beta) \exp\{\lambda^T T(\beta)\} }{\int \exp \{ \lambda^T T(\beta)\} H(\beta) d \beta} \frac{\int \exp \{ \lambda^T T(\beta)\} H(\beta) d \beta}{\int H(\beta) \exp\{\lambda^T T(\beta)\} d\beta} d\beta = \\ & \int T(\beta) \frac{H(\beta) \exp\{\lambda^T T(\beta)\} }{\int \exp \{ \lambda^T T(\beta)\} H(\beta) d \beta}  d\beta = \int T(\beta) q(\beta) d\beta = \mathbb{E}_{q(\beta)}[T(\beta)].
\end{align}

\noindent Now that we have proved \textbf{Lemma H.1}, we know that $\mathbb{E}_{q(\beta)}[ T(\beta)] = \nabla_{\lambda} A_g(\lambda)$; thus, we can write the ELBO as

\begin{align}
\begin{split}
\mathcal{L}(\lambda) &= \mathbb{E}_{-q(\beta)}\big [ \eta_g(x, Z, \alpha)^T \mathbb{E}_{q(\beta)}[ T(\beta) ]\big ] - \lambda^T \mathbb{E}_{q(\beta)}[ T(\beta)] - A_g(\lambda)\\ & = \nabla_{\lambda} A_g(\lambda) \bigg(\mathbb{E}_{-q(\beta)}[\eta_g(x, Z, \alpha)] - \lambda \bigg) - A_g(\lambda).
\end{split}
\end{align}

\noindent Taking the derivative of the ELBO w.r.t. $\lambda$, we get the 

\begin{align}
\begin{split}
\nabla_{\lambda} \mathcal{L}(\lambda) &=  \nabla^2_{\lambda} A_g(\lambda) \bigg(\mathbb{E}_{-q(\beta)}[\eta_g(x, Z, \alpha)] - \lambda \bigg) + \nabla_{\lambda} A_g(\lambda) \bigg( -1 \bigg) + \nabla_{\lambda} A_g(\lambda) \\ &= \nabla^2_{\lambda} A_g(\lambda) \bigg(\mathbb{E}_{-q(\beta)}[\eta_g(x, Z, \alpha)] - \lambda \bigg).
\end{split}
\end{align}

\noindent Knowing that $\nabla^2_{\lambda} A_g(\lambda) = FIM(\lambda)$,\footnote{Fisher Information Matrix.} if we put the above gradient into the formula of the natural gradient, we get 

\begin{equation}
\begin{aligned}
\lambda^{t+1} &= \lambda^t + \gamma \bigg(FIM^{-1}(\lambda) \nabla_{\lambda}\mathcal{L}(\lambda) \bigg) \\
&= \lambda^t + \gamma \bigg( \mathbb{E}_{-q(\beta)}[\eta_g(x, Z, \alpha)] - \lambda^t \bigg) \\
&= (1- \gamma) \lambda^t + \gamma  \mathbb{E}_{-q(\beta)}[\eta_g(x, Z, \alpha)].
\end{aligned}
\end{equation}

\noindent \textbf{Part 2}: We prove that \textbf{Part 1} holds for VaiPhy, i.e., we prove the two corresponding distributions in VaiPhy's ELBO follow EF. That is $q(Z|X)$ and $p(Z, X| \mathcal{B}, \tau)$. Note that if the latter probability  belongs to EF, it implies that $p(Z | X, \mathcal{B}, \tau)$ and $p(X | \mathcal{B}, \tau)$ belong to EF since we know that the likelihood $p(X | \mathcal{B}, \tau)$ is a JC model and itself follows EF; more details of JC being formulated as EF will be provided in the rest of this section.

\noindent In the normal scale, the update equation for latent nodes in the tree is

\begin{equation}
\begin{aligned}
    q^*(Z_u|X) 
    &\propto \prod_{\tau, \mathcal{B}}  \Bigg[ p_\theta(Z_r | \tau)^{I(Z_u=Z_r)} \times p_\theta(Z_u | Y_{pa(u)}, \mathcal{B}, \tau)^{q(Y_{pa(u)}|X)} \\
    &\quad\quad\quad \times \; \prod_{Y_w \in Y_{C_\tau(u)}}  p_\theta(Y_w | Z_u, \mathcal{B}, \tau)^{q(Y_w|X)} \Bigg]^{q(\tau, \mathcal{B} | X)}.
    \label{gzu}
\end{aligned} 
\end{equation}

\noindent We now prove that $q(Z|X)$ follows EF. In Eq. \ref{gzu}, each term involves $p$ with some exponent value calculated by either the previous iteration or the current one---we refer to the exponent as $\alpha$. Each term can be written as a canonical form of EF: $1 \times \exp\{ \alpha \log p\ - 0\}$. Now, $q(Z_u|X)$ is the product over EF distributions; hence $q(Z_u|X)$ also belongs to EF, i.e., the canonical product of two EF is 
\begin{equation}
    h(x) e^{\theta_1 T(x) - A(\theta_1)} \times h(x) e^{\theta_2 T(x) - A(\theta_2)} = \hat{h(x)} e^{(\theta_1 + \theta_2)T(x) - \hat{A}(\theta_1,\theta_2)},
\end{equation} where $\theta$'s are the natural parameters.

\noindent Moreover, in Eq. \ref{gzu}, each $p$, except the probability of the root which is constant, i.e., $0.25^{M}$, follows the JC model's distribution, which is defined by the following, in which up to a constant, belongs to EF.

\begin{align*}
&p(Z_u = i | Z_w = j, \tau, \mathcal{B}) = \frac{1}{4} + \frac{3}{4} e^{-\frac{4}{3} \mathcal{B}_{uw}} \quad \textnormal{if} \quad i=j\\
&p(Z_u = i | Z_w = j, \tau, \mathcal{B}) = \frac{1}{4} - \frac{1}{4} e^{-\frac{4}{3} \mathcal{B}_{uw}} \quad \textnormal{if} \quad i \neq j 
\end{align*}

\noindent To prove that the above belongs to EF, one can rewrite $p(Z_u = i | Z_w = j, \tau, \mathcal{B}) \equptoadd \frac{3}{4} e^{-\frac{4}{3} \mathcal{B}_{uw}}$, where $h(.)=\frac{3}{4}$, $T(.) = -\frac{4}{3}$, and $A(.) = 0$ in the EF's canonical form.
\newline

Next, we prove that $p(Z, X| \mathcal{B}, \tau)$ follows EF. We can refer to all vertices as $Y$, and we again assume the root vertex has a fixed probability that is $0.25^{M}$; therefore, we can write: $p(Y| \mathcal{B}, \tau) = \prod_{i \in V(\tau), i-1=Pa(i)} p(Y_i | Y_{i-1}) p(Y_r | \tau) \propto \prod_{i \in V(\tau), i-1=Pa(i)} p(Y_i | Y_{i-1})$. Now each $p(Y_i | Y_{i-1})$ follows a JC model, and in the previous proof, we showed that JC follows EF. Thus, the product of JC distributions also follows EF, again proven in the previous proof.

\end{document}